\documentclass[sigconf]{acmart}
\AtBeginDocument{%
  }

\copyrightyear{2025}
\acmYear{2025}
\setcopyright{rightsretained}
\acmConference[CHI EA '25]{Extended Abstracts of the CHI Conference on Human Factors in Computing Systems}{April 26-May 1, 2025}{Yokohama, Japan}
\acmBooktitle{Extended Abstracts of the CHI Conference on Human Factors in Computing Systems (CHI EA '25), April 26-May 1, 2025, Yokohama, Japan}\acmDOI{10.1145/3706599.3720050}
\acmISBN{979-8-4007-1395-8/2025/04}

\usepackage{makecell}  
\usepackage{array}
\usepackage{multirow}
\usepackage{float}
\usepackage{caption} 
\usepackage{booktabs}

\begin{document}

\title[Visual Embedding of Screen Sequences for User-Flow Search]
{Visual Embedding of Screen Sequences for User-Flow Search in Example-driven Communication}

\author{Daeheon Jeong}
\authornote{Both authors contributed equally to this research.}
\orcid{0009-0001-2226-5734}
\affiliation{%
  \institution{School of Computing, KAIST}
  \city{Daejeon}
  \country{Republic of Korea}}
\email{daeheon.jeong@kaist.ac.kr}

\author{Hyehyun Chu}
\authornotemark[1]
\orcid{0009-0006-0256-5277} 
\affiliation{%
  \institution{School of Computing, KAIST}
  \city{Daejeon}
  \country{Republic of Korea}}
\email{hyenchu@kaist.ac.kr}

\renewcommand{\shortauthors} {Jeong, Chu, et al.}

\begin{abstract}

Effective communication of UX considerations to stakeholders (e.g., designers and developers) is a critical challenge for UX practitioners. 
To explore this problem, we interviewed four UX practitioners about their communication challenges and strategies.
Our study identifies that providing an example user flow—a screen sequence representing a semantic task—as evidence reinforces communication, yet finding relevant examples remains challenging.
To address this, we propose a method to systematically retrieve user flows using semantic embedding.
Specifically, we design a model that learns to associate screens' visual features with user flow descriptions through contrastive learning.
A survey confirms that our approach retrieves user flows better aligned with human perceptions of relevance.
We analyze the results and discuss implications for the computational representation of user flows.
\end{abstract}



\begin{teaserfigure}
  \includegraphics[width=\textwidth]{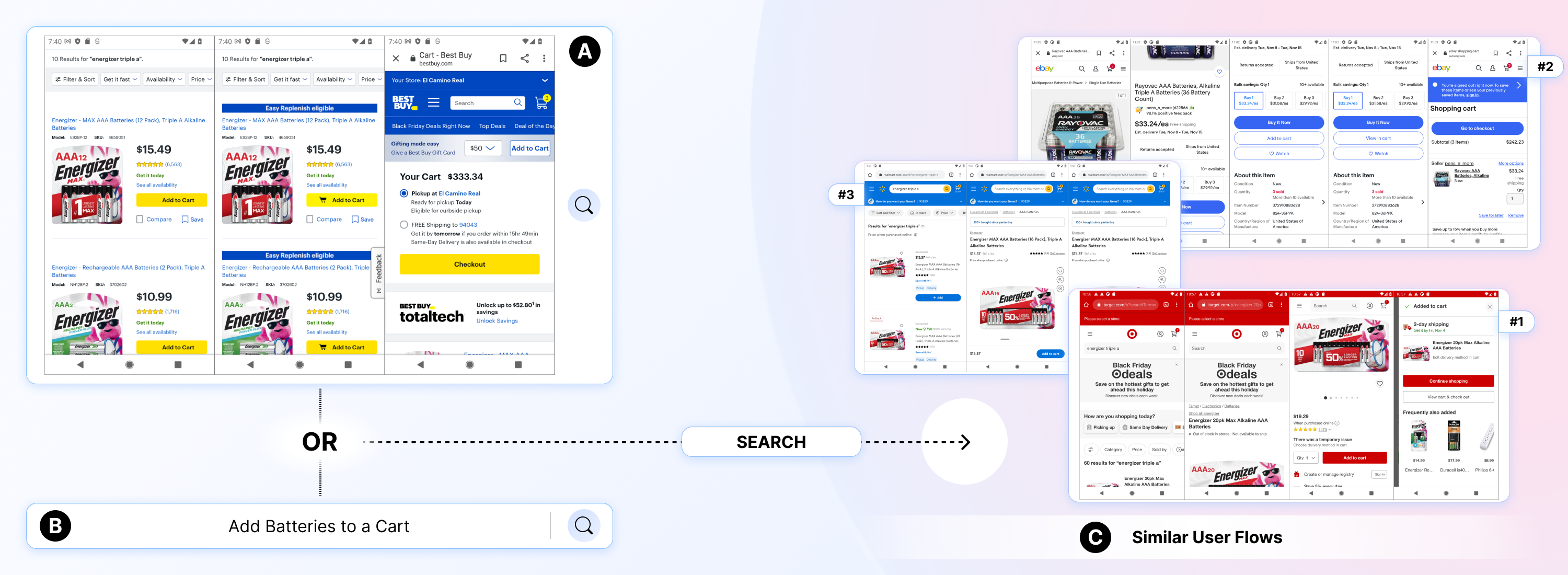}
  \caption{An example retrieval results from our embedding model. Given (A) user flow screen images or (B) textual descriptions, (C) the model retrieves semantically relevant user flows as screen sequences. The numbers attached to each screen sequence indicate its similarity rank to the query.}
  \Description{An example retrieval results from our embedding model. The figure shows the two input methods: a screen sequence and text query "Add batteries to a cart" and their relevant retrieval result from user flow database based on the semantic similarity. Specifically, the figure consists of three user flow examples and their similarity ranks.}
  \label{fig:teaser}
\end{teaserfigure}


\maketitle

\section{Introduction}
Effective communication of user experience (UX) consideration to stakeholders is critical to user-centric product design~\cite{shukla2024communication}.
UX practice is a collaborative process involving stakeholders with diverse professions (e.g., product management, design, development)~\cite{feng2023understanding}.
Accordingly, communicating the importance of UX considerations to other stakeholders is a critical challenge~\cite{norgaard2006usability, gulliksen2006usability, fan2020practices}.
The key characteristic of the challenge is to maximize the buy-in of the stakeholders to make user-aligned decisions by strengthening the argument~\cite{macdonald2022and}.
In response, previous research highlights the communication strategies including rhetoric techniques~\cite{dumas2004describing, gray2016s, rose2016arguing}, exposing stakeholders to a user~\cite{folstad2012analysis}, and visual illustrations~\cite{hornbaek2005comparing, rose2016arguing, gray2016s, macdonald2022and} to demonstrate the significance of the UX considerations.

Among these approaches, example-driven communication suggests effectiveness~\cite{karlgren2012use}, allowing stakeholders an empirical observation of considerations~\cite{herring2009getting, kang2018paragon}. 
However, UX practitioners report difficulties searching for examples precisely aligned with their considerations~\cite{wu2021exploring, kihoon2024genquery}. 
Prior research addresses this challenge with machine learning models that embed information from single-screen examples, leveraging view hierarchies, spatial layouts~\cite{deka2017rico, jiang2024graph4gui, huang2019swire}, and screen text or graphics~\cite{bai2021uibert, liu2018learning, bunian2021vins} for semantic search. 
Although these approaches capture screen-level semantics, a single screen cannot illustrate the user flow, a screen sequence that represents a semantically meaningful task~\cite{kang2018paragon, deka2016erica}.
Information on user flow is key to examples as they demonstrate the design rationales through user interaction scenarios.
Yet, only a few studies have addressed embedding user flows via screen sequences~\cite{deka2016erica, li2021screen2vec, he2021actionbert}, as the problem requires the model to learn both the visual and temporal dimensions of the screens and associate them.


Building on these observations, we propose a method to systematically search user flow examples for evidence-driven communication. 
To ground our approach, we conducted a focus group with four UX practitioners to explore communication strategies. 
Our findings confirm that practitioners use concrete user flow examples to communicate UX considerations, especially during the planning phase. 
To support the systematic search of user flow examples, we propose a method for capturing user flow semantics from visual features in a screen sequence.
Inspired by video classification research~\cite{zhao2024videoprism,arnab2021vivit,yuan2023videoglue}, the model combines visual features using multi-head attention pooling and aligns them with user flow descriptions via contrastive learning on screen sequence–text pairs. 
Our model allows semantic search of user flows by extracting embeddings from an input screen sequence (Figure~\ref{fig:teaser}A) or text (Figure~\ref{fig:teaser}B) and retrieving the most relevant examples from a database using cosine similarity (Figure~\ref{fig:teaser}C). 
A user survey comparing our approach with retrieval based on text embeddings shows that our method aligns more closely with human perceptions of relevance.

In summary, this paper contributes (1) an in-depth interview study on how UX practitioners communicate UX considerations to stakeholders, (2) a novel embedding method for user flow for semantic search, and (3) a survey study that evaluates the effectiveness of this approach and provides implication for computational representation of user flow.
\section{Related Works}
\subsection{Example-driven Communication of UX Considerations}

Communication of UX considerations to stakeholders is critical in collaboration for user-centric product design \cite{gray2016s, rose2016arguing}, helping product managers and developers embrace user-centric decisions \cite{shukla2024communication, macdonald2022and}.
Effective communication addresses two major collaboration challenges: keeping pace with collaborators and aligning priority conflicts \cite{macdonald2022and}.
It ensures the UX is sufficiently considered among collaborators in the fast-paced product development process \cite{holtzblatt2014communicating}.
Also, it aligns UX considerations with other collaborators' priorities (e.g., retention and conversion metric) \cite{shukla2024communication}.
To support communication, previous research explored the strategies including rhetoric techniques \cite{rose2016arguing, dumas2004describing, gray2016s}, exposing stakeholders to a user \cite{folstad2012analysis}, and visual illustrations \cite{hornbaek2005comparing, rose2016arguing, gray2016s, macdonald2022and} to demonstrate the significance of the UX considerations.

Example-driven communication is one of the effective communication strategies~\cite {karlgren2012use}.
In design practices, examples inspire practitioners by showing how content structures and integrates~\cite{lee2010designing, swearngin2018rewire, choi2024creativeconnect}. 
Similarly, when collaborating with stakeholders, examples strengthen the communication by allowing empirical observation~\cite{herring2009getting, kang2018paragon}.
They clarify the rationale behind considerations \cite{kang2018paragon} and help establish a shared understanding of problems \cite{karlgren2012use}. 
While examples take various forms, a prominent theme is user flow, incorporating journey maps, video/audio reels, flows, and user stories \cite{macdonald2022and}. 
User flow represents semantics that arise from interaction sequences such as user task~\cite{deka2016erica}, which is more challenging to infer from than static screens.
In response, this work addresses the computational method to help practitioners search user flow examples through semantic embedding.

\subsection{Semantic Embedding of Screens}

Semantic embedding of screens, which produces computational representations received continuous focus for downstream tasks such as search~\cite{huang2019swire}, screen captioning~\cite{wang2021screen2words}, component prediction~\cite{li2021screen2vec, jiang2024graph4gui}, and code generation~\cite{moran2018machine}. 
A mainstream approach leverages view hierarchy and layout information as representative features~\cite{deka2017rico}, demonstrating embedding models' ability to identify visually coherent examples~\cite{deka2017rico, huang2019swire}. 
Studies further add semantic tags for each layout components~\cite{li2021screen2vec, bunian2021vins, liu2018learning, wu2021screen} or text and images with pre-trained encoders~\cite{wang2021screen2words, ang2022learning, li2021vut, baechler2024screenai, bai2021uibert}, effectively capturing domain-specific screen characteristics like app category or feature~\cite{wang2021screen2words, ang2022learning, li2021vut, bai2021uibert}.
Especially, Wu et al. incorporated temporal dimension in embedding to capture the screen animation~\cite{wu2020predicting}

Furthermore, studies explored methods to embed user flows through screen sequences, which represent a meaningful task through step-by-step interactions~\cite{deka2016erica}.  
Embedding screen sequences expands computational capabilities including user flow search~\cite{deka2016erica}, screen relation inference~\cite{he2021actionbert, li2021screen2vec}, and interaction prediction~\cite{rawles2024androidinthewild}. 
The primary approach involves learning temporal context with order-sensitive models~\cite{he2021actionbert, li2021screen2vec}, and studies explored large language models' capability to comprehend user flows~\cite{lu2024flowy, chen2024gui}. 
Recently, Yicheng et al. proposed a JEPA-based design~\cite{assran2023self} that generates user flow video embeddings to represent user intent~\cite{fu2024ui}. 
Building on recent video understanding models~\cite{zhao2024videoprism, arnab2021vivit, yuan2023videoglue}, we address this problem with order-sensitive multi-head attention pooling of visual features in screen sequences.
\section{Formative Study}
We conducted a formative study with UX practitioners to investigate the challenges of example-based communication. 
Specifically, we aimed to narrow the understanding by exploring two key questions: (Q1) In what contexts does example-driven communication arise? and (Q2) What attributes of examples are most important to practitioners? 
To develop a generalizable understanding of the problem, we conducted a focus group interview~\cite{wilkinson1998focus}.

\subsection{Study Design}

We recruited four UX practitioners with professional experience through snowball sampling. 
The group consisted of three product designers (N = 3) and one UX researcher (N = 1), with an average of 1.27 years (SD = 7.8 months) of professional experience. 
To encourage detailed responses, the group interview was conducted in the interviewees' primary language.
Two researchers collaboratively analyzed the study data through thematic analysis \cite{saldana2021coding}.

The study began with a five-minute introductory session of the study protocol and consent.
The main interview was divided into two sub-sessions, each lasting around 40 minutes. 
In the first sub-session, we focused on (Q1) identifying the context of example-driven communication.
We asked questions about interviewees' current collaboration patterns, existing communication challenges, and strategies they used to address these challenges. 
In the second session, we focused on (Q2) understanding the important attributes of examples.
We introduced a hypothetical AI support tool scenario that provided examples to interviewees and asked them for evaluation.
The researchers ended the interview by summarizing the discussion and asking for additional feedback.

\subsection{Findings}

\subsubsection{Moderating Conflict Through Evidence}

Participants noted that moderating conflict within the organization is a key communication challenge.
With constantly shifting collaborators (P1, P2, P3), participants reported experiencing various conflicts over UX considerations. 
The major source of conflict was each team's different priorities.
These priorities influenced the UX considerations addressed (P2, P3) and their relative importance in the schedule (P3, P4). 
The conflict over priorities stems from different estimations of user experience impact; P4 stated ``\textit{For instance, while some consider increasing the purchase conversion rate most important, others might view app inflow rate as more critical, emphasizing the overall main UI as more important.}'' 

In response, all participants emphasized the importance of evidence in moderating the conflict.
Evidence helps persuade stakeholders regarding UX considerations (P1, P3, P4) and accelerates decision-making (P3). 
User research data, examples, design artifacts, professional experience, and academic research are major types of evidence. 
Participants also combined different evidence types to strengthen communication, e.g., comparing user data between their product and the example (P1) or matching user feedback with example cases (P3). 

\subsubsection{Characteristics of Examples as Evidence}

Participants commonly utilized examples as evidence during the planning phase to address uncertainty.
In the planning phase, the lack of user data and artifacts introduced uncertainty to communication (P1, P3, P4). 
To address the problem, participants searched for existing examples as evidence (P3, P4). 
Upon searching, they collected examples directly related to the product domain, such as examples from main competitors (P2, P3, P4). 
They looked for relatable aspects from examples to the current product; P2 noted, ``\textit{By looking at the examples from competitors, we can see how their approach applies to our problem and identify the relevant issues or pain points.}'' 

In this context, participants viewed examples as more abstract than a component or screen, relating to a user flow.
Participants described examples in words such as product (P4), pattern (P1), and feature (P2, P3). 
Correspondingly, they used modalities that effectively represent user flow, including interaction logs (P1), screen recordings (P2), and flow maps (P3) to communicate the examples to stakeholders. 
These examples allowed stakeholders to make empirical observations on the UX considerations and explore solution scenarios (P1, P2, P3). 
\section{Methods}
\begin{figure*}[ht]
    \centering
    \includegraphics[width=\textwidth]{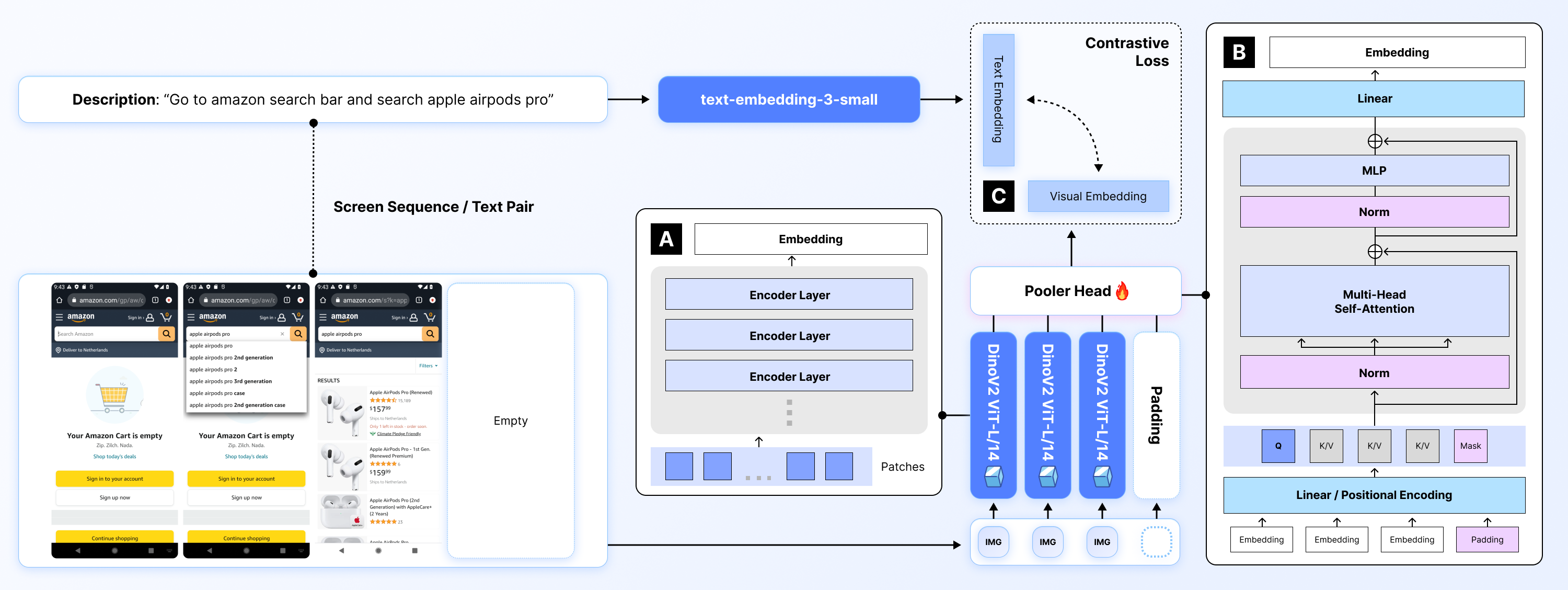} 
    \caption{Our method consists of three main components: (A) representing screen sequences as a series of visual features using ViT; (B) aggregating variable-length visual features using one-layer multi-head attention pooling with temporal encoding and masking; and (C) training the model on contrastive loss between screen sequence–text pairs.}
    \Description{In the figure, a sequence of mobile app screens is processed by a Vision Transformer (ViT) to produce patch-based visual embeddings, which are then aggregated with a single-layer multi-head attention mechanism using temporal encoding and masking. Meanwhile, the text instruction (“Go to Amazon search bar and search Apple AirPods pro”) is transformed by a text embedding model. Finally, a contrastive loss aligns the aggregated screen sequence representation with its corresponding text embedding, allowing the model to learn how screen sequences and text instructions match.}
    \label{fig:architecture}
\end{figure*}

In this section, we present our approach for embedding screen sequences to represent user flows.
User flow is a screen sequence that represents a semantic task~\cite{kang2018paragon, deka2016erica}.
Drawing inspiration from recent video classification models~\cite{zhao2024videoprism, arnab2021vivit, yuan2023videoglue}, we design a model that pools visual features from screen sequence to align them with the user flow description using contrastive learning. 

\subsection{Model Architecture}
To enable semantic embedding of screen sequences, we design an encoder-pooler head structure outlined by Yuan et al.~\cite{yuan2023videoglue} in which the pooler aggregates information from encoders. 
The pooler head is a single cross-attention layer, capable of attending to meaningful visual features through a single learnable query token and self-attention.
Our model captures the key visual features relevant to user flow-related semantics of the screen sequence. 
Additionally, we add padding and masking in the pooler to allow the embedding of variable-length screen sequences.

\textbf{Visual Encoder.} (Figure~\ref{fig:architecture}A) To capture visual features from each screen, we employ DinoV2 (ViT-L/14) with registers~\cite{darcet2023vision,oquab2023dinov2} as a frozen image encoder that takes 224$\times$224 screen images and outputs 1024-dimensional feature embeddings.
Although end-to-end fine-tuning improves the performance \cite{yuan2023videoglue}, we assume that ViT can capture the screen semantics without fine-tuning based on an observation with CLIP~\cite{radford2021learning} by Seokhyeon et al.~\cite{park2023computational}.

\textbf{Pooler Head.} (Figure~\ref{fig:architecture}B) 
After the visual encoder extracts per-screen embeddings, we feed each sequence of embeddings into a learnable pooler head that produces a single vector representation.
When feeding in, to handle variable-length sequences, we pad shorter embedding arrays to a uniform length and use masking to make the pooler focus only on valid frames.
We base our pooler head on a \emph{single-layer} multi-head attention pooling mechanism~\cite{yuan2023videoglue}. 
The model first performs linear projection of each 1024-dimensional screen embedding into a 256-dimensional space and adds a learnable positional embedding to reflect the temporal order. 
Next, a single query token attends to all positions in the sequence via multi-head attention (MHA). 
The input and aggregated output are normalized, passed through a small multi-layer perception (MLP), and finally projected to 1536 dimensions to match the size of text embeddings for user flow description. 
This single-query design allows the pooler head to understand the underlying spatial-temporal relationship between screens in a screen flow.

\subsection{Dataset}
We train our model using data from the Android in the Wild (ATIW)~\cite{rawles2024androidinthewild}, which provides extensive sequences of real-world device interactions. 
In particular, we focus on a subset of the ``SINGLE'' partition, which focuses on the single-user flows. 
While keeping the variable length of each sequence, we filter sequences to those with 3--6 screens each, resulting in 12{,}659 final sequences (episodes) and a total of 49{,}590 screens. 
Our motivation is to isolate a moderate range of screen lengths often encountered in single-user flows (e.g., a few steps to add an item to a cart) and reduce the variability within the dataset. 
We did not include the visual interaction traces between screens in the image.
Each screen image is resized to 224$\times$224 pixels during preprocessing through non-uniform scaling to match the input constraints of DinoV2~\cite{oquab2023dinov2}.
Additionally, we used OpenAI’s \texttt{text-embedding-3-small}~\cite{openai@text}, to generate 1536-dimensional embeddings for each sequence's task description provided by ATIW.

\subsection{Training Configurations}
Our pooler head is trained to align screen-sequence embeddings with text embeddings through a contrastive learning objective similar to CLIP (Figure~\ref{fig:architecture}C)~\cite{radford2021learning}.
Specifically, we adopt a method suggested by Zhao et al.~\cite{zhao2024videoprism} in which the model minimizes a symmetric cross-entropy loss over the similarity scores of all image sequence-text pairs in a mini-batch:
\begin{equation}
\mathcal{L}_{\mathrm{contrast}} \;=\; \tfrac{1}{2}\;\Big[\mathrm{CE}\big(\mathbf{V}\mathbf{T}^{\mathsf{T}}/\tau\big)\;+\;\mathrm{CE}\big(\mathbf{T}\mathbf{V}^{\mathsf{T}}/\tau\big)\Big],
\end{equation}
where $\mathbf{V}$ and $\mathbf{T}$ denote mini-batch matrices of normalized screen-sequence and text embeddings, and $\tau=0.07$ is a fixed temperature. 
For training, randomly divide our dataset into training (90\%) and validation (10\%) splits, and train the model for 100 epochs using the Adam optimizer at a learning rate of $1\times10^{-4}$, with a batch size of 1024 episodes per iteration. 
During each gradient update, the pooler attends the visual embedding of screens and produces a single output, which is then compared against the corresponding textual embedding via the contrastive loss. 
Since the visual and text encoder remain frozen, the network converges quickly.
We observed that the validation loss stabilized at approximately 100 epochs with a value of 2.616 (using PyTorch seed 123).
\section{Evaluation}
We conducted a user study to evaluate the performance of our model and investigate its practical applications in design processes.
In the study, our objective was to address two key questions: (1) how our approach compares to the text description-based baseline and (2) how it could be integrated into real-world design practices.
Additionally, we report a minor implementation error in the model; the impact on performance was minimal.

\subsection{Study Design}

We recruited 21 participants—14 with UX experience and 7 without—for an online survey that included 27 multiple-choice questions and 15 optional open-ended questions.
The study consisted of two main tasks.

In \textbf{Task 1}, participants assessed the similarity between a source screen sequence and two candidate sequences; each candidate was obtained using the baseline model and our model.
We randomly sampled five source screen sequences.
We measured the similarity using a 5-point Likert scale across four dimensions (service similarity, screen type similarity, content similarity, and visual similarity).
These dimensions reflect established practices for assessing screen-based relevance and layout coherence~\cite{li2021screen2vec}.
To minimize potential order bias, we randomized the presentation order of two candidate sequences.
In \textbf{Task 2}, participants assessed design examples retrieved by our model in practical scenarios (e.g., "designing a search result page") from a UX practitioner's perspective.
Based on established design applicability measures~\cite{jeon2021fashionq}, participants rated each candidate example on five aspects using a 5-point Likert scale: layout alignment, UI component utility, innovation potential, integration feasibility, and overall reference value.

For data analysis, we examined the responses of both tasks using different analytical approaches.
For Task 1, we verified the normality of the data distribution using the Shapiro-Wilk test (\( p > 0.05 \)), confirming normality, and subsequently conducted paired t-tests to statistically compare our approach with the baseline.
For Task 2, we employed a descriptive analysis of the quantitative Likert-scale response.

\subsection{Findings}

\subsubsection{Effectiveness of Sequence-based Search}

\begin{figure*}[h]
    \centering
    \includegraphics[width=\textwidth]{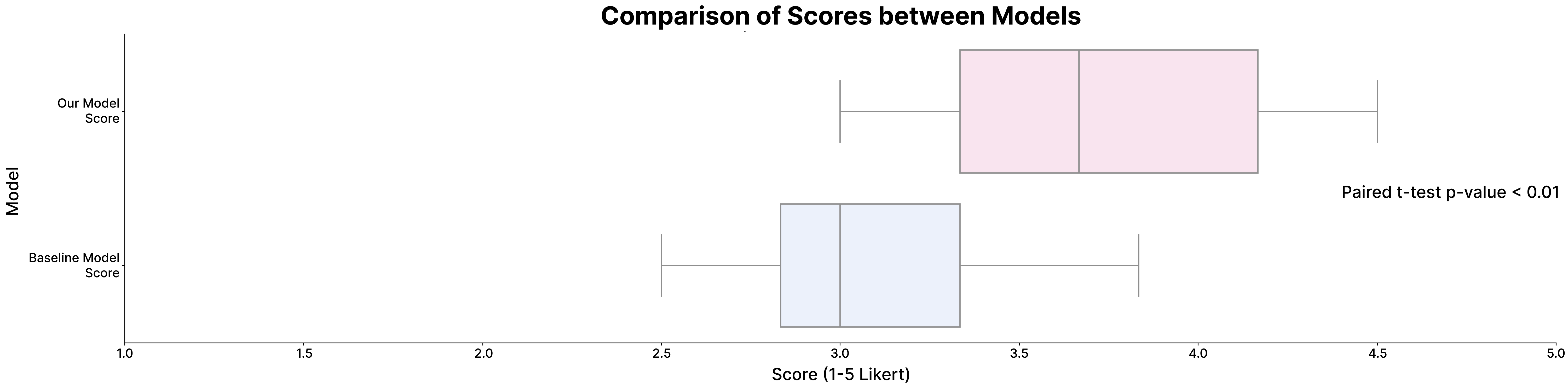}
    \caption{Box plots comparing similarity scores between our sequence-based search model and the baseline model (\( n = 21 \)). The plot illustrates median scores, quartiles, and individual data points for both models. A paired t-test revealed significant differences between the models (\( p < 0.01 \)).}
    \Description{The figure presents two box plots comparing scores (on a 1–5 Likert scale) between the baseline model (lower scores) and our sequence-based search model (higher scores). Each box shows the median, quartiles. The difference between the two models’ scores is statistically significant.}
    \label{fig:model-comparison}
\end{figure*}

As shown in Figure~\ref{fig:model-comparison}, our sequence-based search model demonstrated statistically significant higher similarity scores compared to the baseline condition (\( t(20) = 8.50, p < .001 \)).
The high \( t \)-statistic (8.50) indicates not just statistical significance but also a substantial practical difference in performance.
This improvement was observed consistently across participants with UX experience (\( n = 14 \)) and those without (\( n = 7 \)).

The difference between our model and the baseline was modest in magnitude (\( M = 0.61, SD = 0.32\)) but consistent.
A detailed analysis revealed that participants with UX experience rated our model higher (\( M = 3.83 \)) than the baseline (\( M = 3.20 \)).
Similarly, participants without UX experience also preferred our model (\( M = 3.48 \)) over the baseline (\( M = 2.88 \)).
Notably, both groups showed comparable levels of improvement (UX experienced: \( +0.63 \), non-experienced: \( +0.60 \)), suggesting the model's effectiveness generalizes across different levels of UX expertise.

\subsubsection{Practical Value as UX Design Reference}

The model's value as a design reference was evaluated across five key dimensions, with all aspects scoring above 3.50 on average (on a 5-point scale). 
Participants positively assessed the system's overall reference value for design work (\( M = 3.67, SD = 1.16 \)).
One UX designer particularly emphasized the benefit of viewing consistent UI sequences together, noting, ``\textit{it's nice to see the sequence of consistent UI because, if I search them separately, that could create inconsistencies in my UI flow.}''

Participants rated the system highly on ease of integration into their current design processes (\( M = 3.83, SD = 1.19 \)).
Our model also received favorable evaluations regarding the practicality of the UI component (\( M = 3.68, SD = 1.09 \)) and the applicability of the layout and user flow (\( M = 3.63, SD = 1.11 \)).
Correlation analysis indicated a strong relationship between UI component practicality and layout applicability (\( r = 0.86 \)).
Several participants highlighted the integrated value of layouts and UI components, as reflected in comments like, ``\textit{I think the structural and button designs could be good references for UI development.}''

\subsubsection{Design Convention and Innovation}

The innovative design solutions aspect received relatively lower ratings (\( M = 3.24, SD = 1.16 \)) and showed weaker correlations with other evaluation aspects (\( r = 0.49 \text{--} 0.74 \)).
Participants' responses indicated a preference for conventional UI practices.
One UX practitioner noted that ``\textit{these are common UI practices,}'' highlighting the value of capturing established UI conventions.
Another participant commented, ``\textit{Why does the UI need to be newly designed or innovative in the first place? Doesn't that make the user flow more unfamiliar?}''

\section{Discussion}
\subsection{Reinforcing User Flow Semantics with Visual Information}

The evaluation result suggests that visual information can effectively reinforce user flow semantics. 
Our model achieved statistically higher ratings consistently, indicating a stronger alignment with human perception of similarity. 
Notably, despite the statistical significance and consistent gap, the average rating difference between our model and the baseline was relatively moderate.
This indicates that the visual features captured by our model consistently add the information available in the textual description that predicts the semantic similarity between the two sequences.
Specifically, we observed that screen sequences retrieved by our model shared a series of visual features with the source screen sequence, and these features add precision to the task description (e.g., the keyword “search” retrieved screen sequences that contain interactions with a search bar). 
Considering the model’s simple architecture, the model's capability to link textual semantics to a specific sequence of visual features is promising.

\subsection{Design Implications for Sequence-based UI Reference Systems}

Based on our formative study, our sequence-based approach demonstrates the potential to enhance example-driven communication in UX by providing rich user journeys, extending beyond isolated examples \cite{karlgren2012use}.
Regarding this, features that connect sequence examples with journey maps or task flows could enhance contextual understanding \cite{karlgren2012use, macdonald2022and}.
Additionally, improving system explainability would boost trustworthiness as UX communication has an evidence-based nature \cite{herring2009getting, kang2018paragon}.
Furthermore, for the effective utilization of retrieved sequences, careful integration with existing design tools such as Figma or Sketch would be recommended \cite{figma, sketch}.

Another notable finding was the UX search expectation between convention and innovation.
Our sequence-based search has the feasibility to help identify established UI patterns.
Furthermore, the user preference for familiar UI conventions reflects practical reference needs in professional settings.
However, to support diverse practical scenarios, supporting innovative design exploration for diversity aspects can be considered for additional search options by providing diversity metrics rather than similarity alone.

\subsection{Limitations and Future Work}

Despite these promising results, considerable uncertainties regarding the model's performance and applicability remain. 
Our model is trained on a narrow subset of the ATIW dataset~\cite{rawles2024androidinthewild} with domain-specific examples, hence the ability to handle diverse user flow types remains unknown. 
The model's understanding of textual descriptions is limited to the label distribution in training data, thereby constraining the selection of search queries.
Additionally, the current architecture uses a relatively simple attention-pooling method with a pre-trained ViT and text encoder, focusing primarily on evaluating the model's capabilities. 
To advance this approach, future iterations could incorporate screen text~\cite{wang2021screen2words} and user interactions between screens to model training. 
Incorporating model design strategies such as variable resolution input~\cite{lee2023pix2struct}, low-rank adapters~\cite{yuan2023videoglue}, and masked autoencoders~\cite{tong2022videomae, zhao2024videoprism} could further enhance the performance. 
Integrating language decoders~\cite{li2022spotlight, baechler2024screenai} could extend functionality beyond search to tasks such as interaction localization, question answering, and action prediction with relevant benchmark evaluations.
Similarly, a comparative evaluation with an existing video foundation model~\cite{zhao2024videoprism, wang2022internvideo} and previous screen sequence embedding approaches~\cite{deka2016erica, li2021screen2vec, lu2024flowy} against benchmark can highlight the relative benefit of our approach.
Furthermore, an evaluation with a more diverse UX practitioners can expand exploration towards more specific design scenarios in real-world applications.
\section{Conclusion}
We explored an approach to enhance UX practitioners' ability to communicate design considerations through sequence-based example search.
Our formative study showed that practitioners' use user flow examples to communicate UX considerations.
We developed and evaluated an AI model for embedding screen sequences to extract user flow semantics, which improved human alignment compared to text-based retrieval baselines.
A survey involving 21 practitioners with varying UX backgrounds revealed that our approach outperformed the baseline in retrieving relevant user flow example and showed potential for integration with design practices. 
This research advances the understanding of user flow by exploring its role in communication and its computational representation.

\begin{acks}
We are deeply grateful for our participants and reviewers who significantly contributed to this work.
We especially thank our teammates, Dongyu and Woohyun, for their collaboration throughout the HCI course.  
We also appreciate Jaesang, Hyoungwook, Daehyun, Yeonsu, and Bekzat for providing valuable feedback on our drafts.
\end{acks}

\bibliographystyle{ACM-Reference-Format}
\bibliography{base}

\appendix
\section{Survey Results for Similarity Evaluation in Task 1}

\begin{table}[H]
\centering
\renewcommand{\arraystretch}{1.3}
\begin{tabular}{%
  >{\centering\arraybackslash}m{1.2cm} 
  >{\centering\arraybackslash}m{1.8cm}  
  >{\centering\arraybackslash}m{1.0cm}  
  >{\centering\arraybackslash}m{1.2cm} 
  >{\centering\arraybackslash}m{1.5cm}%
}
\toprule
\shortstack{\textbf{Parti-}\\\textbf{cipant}} & 
\shortstack{\textbf{UX}\\\textbf{Experience}} & 
\shortstack{\textbf{Our}\\\textbf{Model}\\\textbf{Score}} & 
\shortstack{\textbf{Baseline}\\\textbf{Model}\\\textbf{Score}} & 
\shortstack{\textbf{Difference}\\\textbf{(Our -}\\\textbf{Baseline)}} \\
\midrule
A1  & Yes      & 4.33 & 3.17 & 1.17 \\
A2  & Yes      & 3.50 & 3.00 & 0.50 \\
A3  & Yes      & 3.17 & 3.00 & 0.17 \\
A4  & Yes      & 4.33 & 3.17 & 1.17 \\
A5  & No  & 3.33 & 3.00 & 0.33 \\
A6  & No  & 3.00 & 2.50 & 0.50 \\
A7  & Yes      & 4.33 & 3.83 & 0.50 \\
A8  & No  & 4.50 & 3.67 & 0.83 \\
A9  & Yes      & 3.67 & 2.50 & 1.17 \\
A10 & No  & 3.50 & 2.83 & 0.67 \\
A11 & No  & 3.83 & 2.83 & 1.00 \\
A12 & Yes      & 3.33 & 3.33 & 0.00 \\
A13 & No  & 3.00 & 2.50 & 0.50 \\
A14 & Yes      & 4.00 & 3.50 & 0.50 \\
A15 & Yes      & 3.50 & 2.83 & 0.67 \\
A16 & Yes      & 3.33 & 3.17 & 0.17 \\
A17 & Yes      & 4.17 & 3.33 & 0.83 \\
A18 & Yes      & 4.33 & 3.50 & 0.83 \\
A19 & Yes      & 4.00 & 3.50 & 0.50 \\
A20 & No  & 3.17 & 2.83 & 0.33 \\
A21 & Yes      & 3.67 & 3.00 & 0.67 \\
\bottomrule
\end{tabular}
\caption{Survey Results for Similarity Evaluation in Task 1. Participants compared the similarity between two candidate screen sequence sets based on a given source screen sequence across four dimensions: service similarity, screen type similarity, content similarity, and visual similarity. All responses were measured on a 5-point Likert scale, where 1 indicates the lowest score (Completely different) and 5 indicates the highest score (Almost identical). To minimize order bias, the presentation order of baseline and proposed model sequences was randomized.}
\label{tab:task1_results}
\end{table}

\section{Scenario example in Task2}

\begin{figure}[htbp]
    \centering
    \includegraphics[width=\linewidth]{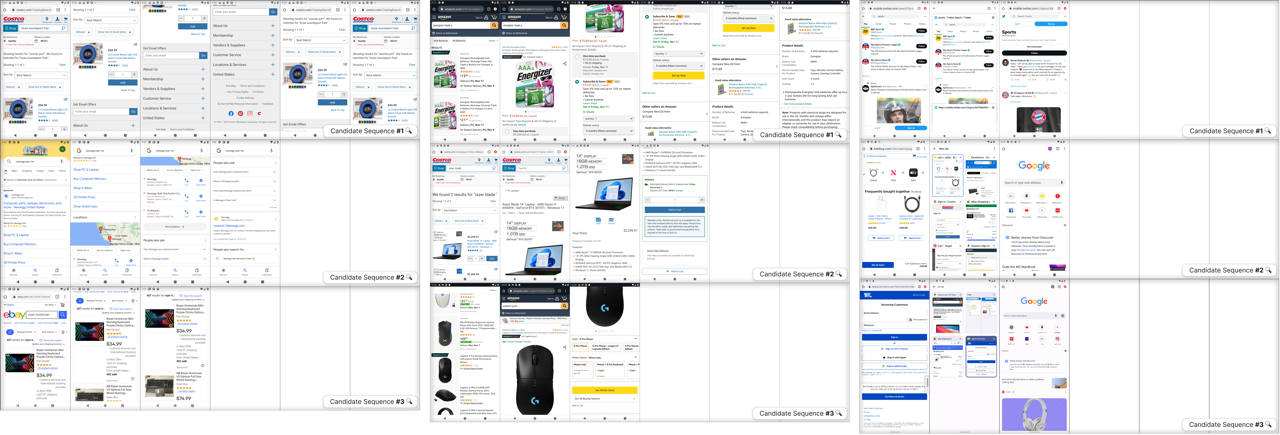}
    \caption{Example search result sequences retrieved for each scenario in Task 2. 
    Scenario 1 focuses on search result pages, Scenario 2 on product detail pages, and Scenario 3 on multi-tab interfaces.}
    \label{fig:task2}
    \Description{This composite image is arranged in a grid with multiple browser screenshots. The first set of screenshots (Scenario 1) shows search result pages listing various products. The second set (Scenario 2) focuses on detailed product pages for items like watches and electronics. The final set (Scenario 3) features a multi-tab interface comparing search results and product details across different sites.}
\end{figure}

This section presents three example scenarios used in Task 2, along with the screen sequences retrieved by our model for each query. Figure~\ref{fig:task2} shows representative results retrieved for each scenario.

\subsection*{Scenario 1}
\begin{description}
  \item[\textbf{\textit{Scenario Description}}]
        You're trying to design a search result page and want to find relevant examples.
    
  \item[\textbf{\textit{Text Query}}]
    Retrieve a pattern, "check a search result"
\end{description}

\subsection*{Scenario 2}
\begin{description}
  \item[\textbf{\textit{Scenario Description}}]
        You're trying to design a product detail page and want to find relevant examples.

  \item[\textbf{\textit{Text Query}}]
        Retrieve a pattern, "scroll to a product detail"
\end{description}

\subsection*{Scenario 3}
\begin{description}
  \item[\textbf{\textit{Scenario Description}}]
        You're trying to design a multi-tab interface and want to find relevant examples.

  \item[\textbf{\textit{Text Query}}]
        Retrieve a pattern, "changing between tabs"
\end{description}

\newpage

\section{Survey Results for Design Process Integration Assessment in Task2}
\captionsetup[table]{width=\textwidth, justification=raggedright, singlelinecheck=false, hypcap=false}
\begin{minipage}{\textwidth}
\centering
\renewcommand{\arraystretch}{1.3}
\begin{tabular*}{\textwidth}{@{\extracolsep{\fill}} p{0.60\textwidth} c c c c c c c}
\toprule
\textbf{Question (5-scale Likert)} & \textbf{Mean} & \textbf{SD} & \textbf{Min} & \textbf{25\%} & \textbf{50\%} & \textbf{75\%} & \textbf{Max} \\
\midrule
This screen's layout and user flow would be applicable to my project & 3.63 & 1.11 & 1 & 3 & 4 & 5 & 5 \\
The UI components would be valuable additions to my design & 3.68 & 1.09 & 1 & 3 & 4 & 4 & 5 \\
This screen offers innovative design solution ideas that I could adapt for my project & 3.24 & 1.16 & 1 & 2 & 3 & 4 & 5 \\
I could easily integrate elements from this screen into my current design process & 3.83 & 1.19 & 1 & 3 & 4 & 5 & 5 \\
Overall, this screen serves as an excellent reference for my design work & 3.67 & 1.16 & 1 & 3 & 4 & 5 & 5 \\
\bottomrule
\end{tabular*}
\captionof{table}{Survey Results for Design Process Integration Assessment. This table summarizes the results of the survey evaluating various aspects of the design process, including layout applicability, value of UI components, innovative solutions, ease of integration, and overall reference value. All responses were measured on a 5-point Likert scale, where 1 indicates the lowest score (Strongly Disagree) and 5 indicates the highest score (Strongly Agree). The metrics include Mean, Standard Deviation (SD), Minimum (Min), 25th Percentile (25\%), Median (50\%), 75th Percentile (75\%), and Maximum (Max).}
\label{tab:task2_results}
\end{minipage}

\end{document}